# Photoluminescence in PbS nanocrystal thin films: Nanocrystal density, film morphology and energy transfer

L. Tsybeskov, M. Alam, S. B. Hafiz, D.-K. Ko, A. M. Bratkovsky, X. Wu, and D. J. Lockwood

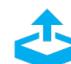

View Online    Export Citation    CrossMark

### ARTICLES YOU MAY BE INTERESTED IN

Efficient green InP quantum dot light-emitting diodes using suitable organic electron-transporting materials
Applied Physics Letters **117**, 111104 (2020); https://doi.org/10.1063/5.0020742

Blue electroluminescent metal halide perovskites
Journal of Applied Physics **128**, 120901 (2020); https://doi.org/10.1063/5.0016377

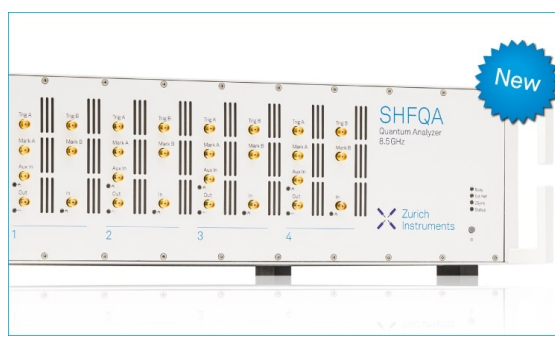

# Photoluminescence in PbS nanocrystal thin films: Nanocrystal density, film morphology and energy transfer

L. Tsybeskov,[1,a]  M. Alam,[1]  S. B. Hafiz,[1]  D.-K. Ko,[1]  A. M. Bratkovsky,[2]  X. Wu,[3] and D. J. Lockwood[3]

**AFFILIATIONS**

[1]ECE Department, New Jersey Institute of Technology, Newark, New Jersey 07102, USA
[2]Kapitza Institute for Physical Problems, 117454 Moscow, Russia
[3]National Research Council, Ottawa, Ontario K1A 0R6, Canada

[a]Author to whom correspondence should be addressed: leonid.tsybeskov@njit.edu

**ABSTRACT**

We show that photoluminescence properties of PbS nanocrystal thin films are directly related to film morphology and nanocrystal density. In densely packed PbS nanocrystal films, low-temperature donor-to-acceptor energy transfer is mainly responsible for the photoluminescence spectra narrowing and shift toward longer wavelengths. At elevated temperatures, back energy transfer is proposed to be responsible for an unusual photoluminescence intensity temperature dependence. In thin films with a low PbS nanocrystal density, the energy transfer is suppressed, and the effect is dramatically reduced.

Published under license by AIP Publishing. https://doi.org/10.1063/5.0022549

## I. INTRODUCTION

During the last several decades, semiconductor nanocrystal (NC) fabrication has been advanced to the level of precise materials design and demonstration of functional optoelectronic devices.[1–5] Colloidal NCs are one of the most studied semiconductor NC systems, and PbS NCs with well-controlled sizes and highly tunable optical properties are produced routinely and studied in great detail.[2,6–8] It has been shown that PbS NC optical absorption is directly linked to the NC sizes due to the quantum confinement effect.[1,8–14] However, there are significant discrepancies in the reported photoluminescence (PL) properties of the similar size PbS NCs,[15–20] and it has been reported that thin films comprised of PbS NCs of the same sizes fabricated by drop casting (DC) and spin casting (SC) techniques showed quite different PL spectra and PL intensity temperature dependencies.[20] This is a very surprising observation because, similar to optical absorption, the PL spectra are expected to be determined mainly by the NC size and size distribution. To address this issue, various exciton recombination mechanisms involving NC surface states have been proposed.[18–20] As an alternative explanation, it has been suggested that in densely packed PbS NC films, an efficient energy transfer (ET) process takes place with smaller NCs acting as donors and larger NCs acting as acceptors.[20–23]

In general, in a system comprised of the same type of semiconductor NCs, two major mechanisms of ET should be considered, the Dexter energy transfer (DET), which is based on exchange interactions between donor and acceptor NCs,[24,25] and the Förster resonant energy transfer (FRET), which is a dipole–dipole coupling phenomenon mainly governed by the spectral and spatial overlaps between donor and acceptor NCs. In the original FRET model with a donor interacting with a point-like acceptor, the FRET rate depends on the distance separating the dipoles as $1/d^6$.[24,25] However, in various geometries, the FRET rate could be proportional to $1/d^n$ with $3 < n < 4$, and the Förster radius is reported to be close to 5 nm.[25] Regarding the spectral overlap, in a system with a reasonable NC size distribution, an overlap between the luminescence spectra of donors and absorption spectra of acceptors is expected. Note that the back ET (from donor to acceptor) is possible but only in the case of the relaxed acceptor gaining enough thermal energy; thus, the back ET should be suppressed at low temperatures.

Some of the key results supporting the FRET model are the demonstration that the length of organic ligands separating the NCs controls the ET rate, which is proportional to $1/d^n$,[26] and the experimental proof of an efficient exciton funneling.[27] In addition, time-resolved PL experiments showed a rising PL component, which has been attributed to a cascaded, multistep, from-smaller-to-larger NC FRET mechanism.[21,28] In agreement with this model, it has been shown that under pulsed excitation, the shorter-wavelength (or higher photon energy) PL decays faster compared to the longer-wavelength (or lower photon energy) PL, causing the decaying PL peak to shift progressively toward longer wavelengths.[21,22]

Understanding ET in semiconductor NCs and the design of effective ET donor–acceptor complexes are complicated by several factors including uncontrollable broadening of the NC optical spectra. Even within a perfectly monodisperse NC ensemble, temperature-dependent homogeneous broadening due to exciton–phonon interactions is expected.[29–31] In real NC systems, despite the demonstration of various techniques of the size distribution reduction, the NC size distribution is far from being monodispersed. Thus, a combination of thermal broadening, considerable distribution of NC sizes, uncertainties in the NC shapes, and crystallographic orientation provides additional challenges to the understanding of the ET and exciton recombination mechanisms in NC films. The goal of this paper is to clarify a relationship between the PbS NC thin film density, morphology, and the ET processes.

## II. MEASUREMENTS AND SAMPLES

Transmission electron microscopy (TEM) studies were performed using a JEOL JEM-2100F field emission source transmission electron microscope. Optical absorption measurements were performed at room temperature using a custom UV-Vis-NIR spectrophotometer based on Stellarnet RW-NIRX-SR and BLK-CXR spectrometers. The PL measurements were performed using a 0.5-m focal length Acton Research automated monochromator, a thermo-electrically cooled InGaAs photomultiplier, a 660 nm laser diode for the PL excitation, and a close-cycle optical cryostat with temperature control ranging from 20 to 350 K. The PL measurements were performed using samples deposited on sapphire substrates with surface roughness better than 0.5 nm and thermal conductivity close to 35 W/(m K). At the highest used excitation intensity of ∼100 μW/cm², no sample heating by a laser beam during PL measurements was detected.

PbS NC samples for this work were synthesized using the previously reported fabrication process[13] with the following modifications. Briefly, 0.45 g of lead oxide (PbO), 10 ml of 1-octadecene (ODE), and 2 ml of oleic acid (OLAC) were loaded in a reaction vessel. The vessel was heated to 90 °C under vacuum for 2 h. Afterward, the temperature of the vessel was raised to 165 °C under nitrogen for reaction. A syringe containing 2 ml of bis(trimethylsilyl) sulfide, diluted with 10 ml of anhydrous ODE, was swiftly injected into the vessel. Immediately following the injection, the heating mantle was removed, and the reaction vessel was left to cool down to room temperature in air. The reaction product was purified using hexane and ethanol three times, re-dissolved in tetrachloroethylene (TCE) for optical absorbance measurement or a hexane/octane (10:1 in volume) mixture for film deposition, and kept inside the glovebox. The targeted PbS NC diameter was 3 nm.

Figure 1(a) shows the room temperature optical absorption spectrum of the above described PbS NC colloidal solution with an excitonic peak at ∼1.2 eV. The inset shows a TEM image of a single PbS NC with clearly observed lattice fringes. Figure 1(b) shows the PbS NC size distribution obtained from analyzing TEM images of ∼70 individual NCs. The average NC diameter is close to 3.1 nm with an approximated Gaussian size distribution of ±8%. In these studies, all samples were prepared using only this size PbS NCs.

Samples for TEM measurements were prepared by DC the PbS NC solution on a TEM grid coated with an amorphous carbon film. For sample 1, a solution containing low NC concentration of 2.5 mg/ml was used, yielding a submonolayer film with ∼$9.5 \times 10^{11}$ NCs/cm² area. Figure 2(a) shows a TEM image of a PbS NC sub-monolayer with an 8–12 nm (area dependent) average distance between the NCs. For sample 2, a higher NC concentration (25 mg/ml) was used, resulting in a thicker, close to three monolayers film ($1.6 \times 10^{13}$ NCs/cm² area). Figure 2(b) shows a TEM image of densely packed, partially self-assembled PbS NC films with an average distance between NCs of ∼2 nm.

For PL measurements, the following samples have been used. Sample A was prepared using 2.5 mg/ml NC solution (via a process similar to that for the TEM sample 1). The PL measurements were performed using the area of the film with an anticipated thickness close to a NC monolayer. Sample B was prepared by SC 25 mg/ml NC solution (similar to that for the TEM sample 2). A sapphire substrate was flooded with the NC solution and was spun at 3500 rpm for 3 min. The NC film thickness was highly uniform and estimated to be ∼9.5 nm (i.e., 3 NC monolayers). Sample C was similar to sample A but with a slightly higher NC concentration (10–15 mg/ml).

## III. EXPERIMENTAL RESULTS

Figure 3 compares PL spectra recorded in sample A and sample B at different temperatures. At room temperature, the PL spectra in both samples are almost the same. As the temperature decreases from 300 K to 20 K, the PL spectra shift toward lower photon energy. In sample A, a broad PL spectrum remains featureless within the entire 20 K–300 K temperature range [Fig. 3(a)]. In sample B at temperatures 200 K ≥ T ≥ 100 K, we observe at least two well-defined PL bands [Fig. 3(b)]. At the lowest recorded temperature (T = 20 K), the PL spectra in samples A and B are very different [Fig. 3(c)]: in sample A, the PL peak is at ∼1.03 eV with a full width at half maximum (FWHM) of ∼100 meV, while in sample B, the PL peak is at ∼0.93 eV with FWHM of only 50 meV.

In sample B, for the PL spectra recorded at different temperatures, we perform a multi-peak fitting procedure with background correction[32] and find that two PL band fitting (hereafter denoted as PL1 and PL2 bands) is satisfactory (not shown). In Fig. 4, we present the PL peak position, the PL FWHM, and the PL intensity as a function of temperature for the PL spectrum in sample A and the two PL bands in sample B. In Fig. 4(a), we plot the temperature dependence of the PL peak position for the PL spectrum in sample A, two PL bands in sample B, and the calculated temperature dependence of the energy gap $E_G(T)$ in 3.1 nm diameter PbS nanocrystals. Note that in these calculations, the value of the energy gap for 3.1 nm diameter PbS NCs at T = 300 K is established using optical absorption measurements [Fig. 1(a)], and it is in a good agreement with the results presented in Refs. 14, 33, and 34. The energy gap temperature dependence is calculated using the Varshni equation with parameters described in Refs. 35–37. Figure 4(b) shows the temperature dependence of the PL peak FWHM, and Fig. 4(c) shows normalized integrated PL intensity as a function of temperature (also for the three described PL bands). The experimental results presented in Fig. 4 are summarized below:

1. The PL peak position as function of temperature for all three PL bands correlates well with the calculated energy gap temperature dependence in a 3.1 nm diameter PbS NC. In sample B, the PL Stokes shift notably increases as the temperature decreased below 200 K, while for sample A, the increase is less distinctive [Fig. 4(a)].

2. In sample B, the PL1 and PL2 FWHM temperature dependencies are similar, while the PL2 band is slightly broader [Fig. 4(b)]. In sample A, the PL FWHM temperature dependence is also similar to that in sample B, but the PL spectrum is significantly broader [Fig. 4(b)].

3. Figure 4(c) shows that in sample A, the PL intensity increases as temperature decreases with a maximum at T ≈ 60 K. In sample B, the PL1 band behaves similarly. At the same time, the PL2 band shows the opposite behavior; the intensity decreases as temperature decreases, and the PL2 band practically disappears at T < 60 K.

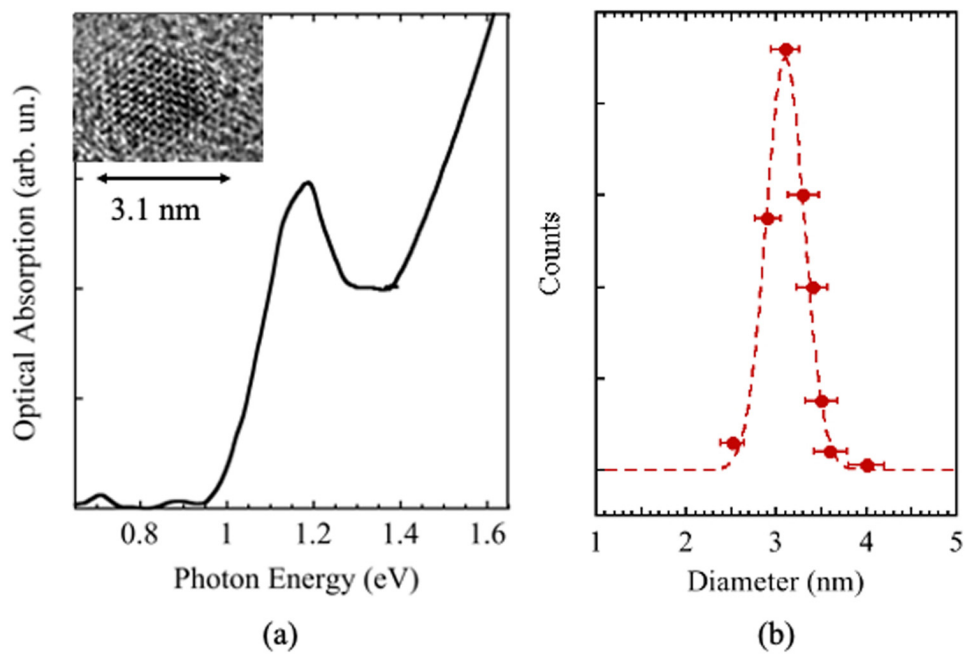

FIG. 1. (a) Optical absorption spectrum and (b) size distribution of PbS nanocrystals (NCs) obtained by analyzing transmission electron microscopy (TEM) data. The inset in (a) shows a TEM image of a 3.1 nm diameter PbS NC with clearly visible lattice fringes.

## IV. DISCUSSION

In sample A with low NC density and an average distance between PbS NCs greater than 8 nm, FRET is expected to be inefficient and could be neglected. As the temperature decreases, the PL spectrum peak position as a function of temperature, in general, follows the calculated temperature dependence of the effective energy gap of a 3.1 nm diameter PbS NC, and the PL Stokes shift slightly increases at T < 150 K (similar to the results in Refs. 38 and 40). Using the calculated dependence of the energy gap as a function of reciprocal PbS NC diameters (based on Refs. 14, 34, and 33), we estimate that in a system with 3.1 ± 0.25 nm PbS NCs [Fig. 1(b)], the PL FWHM is expected to be ∼100 meV [Fig. 5(a)]. We find that this result is very close to our experimental data (Figs. 3 and 4). We conclude that in the case of inefficient FRET, the room temperature PL spectrum reflects the PbS NC size distribution and thermal broadening, and at low temperature, the PL FWHM is entirely controlled by the NC diameter distribution.

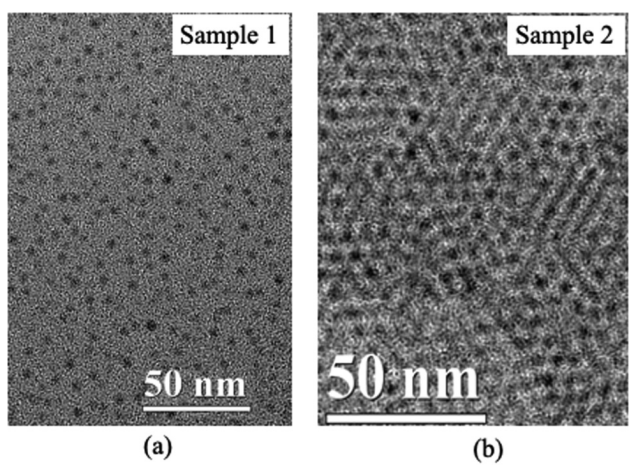

FIG. 2. TEM images of (a) low and (b) high PbS NC density samples.

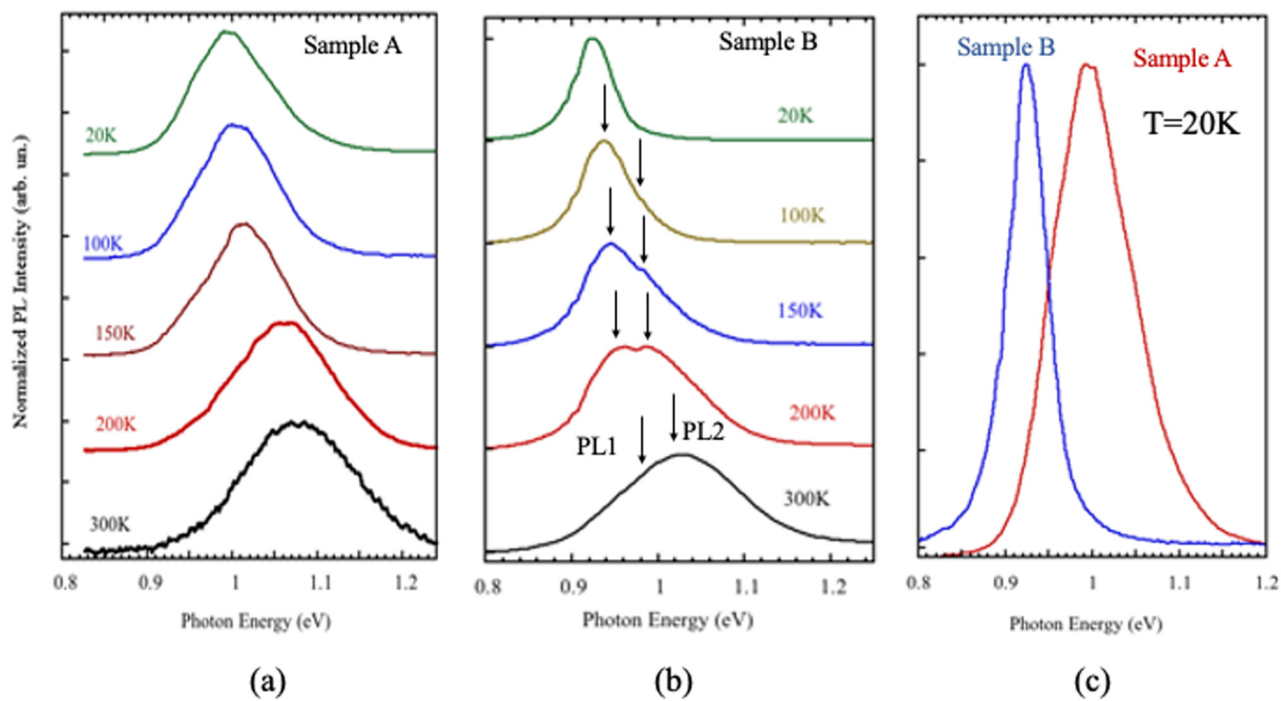

**FIG. 3.** PL spectra recorded at different (indicated) temperatures in (a) sample A and (b) sample B. The PL1 and PL2 bands are shown by arrows. (c) Comparison of normalized low-temperature PL spectra in samples A and B.

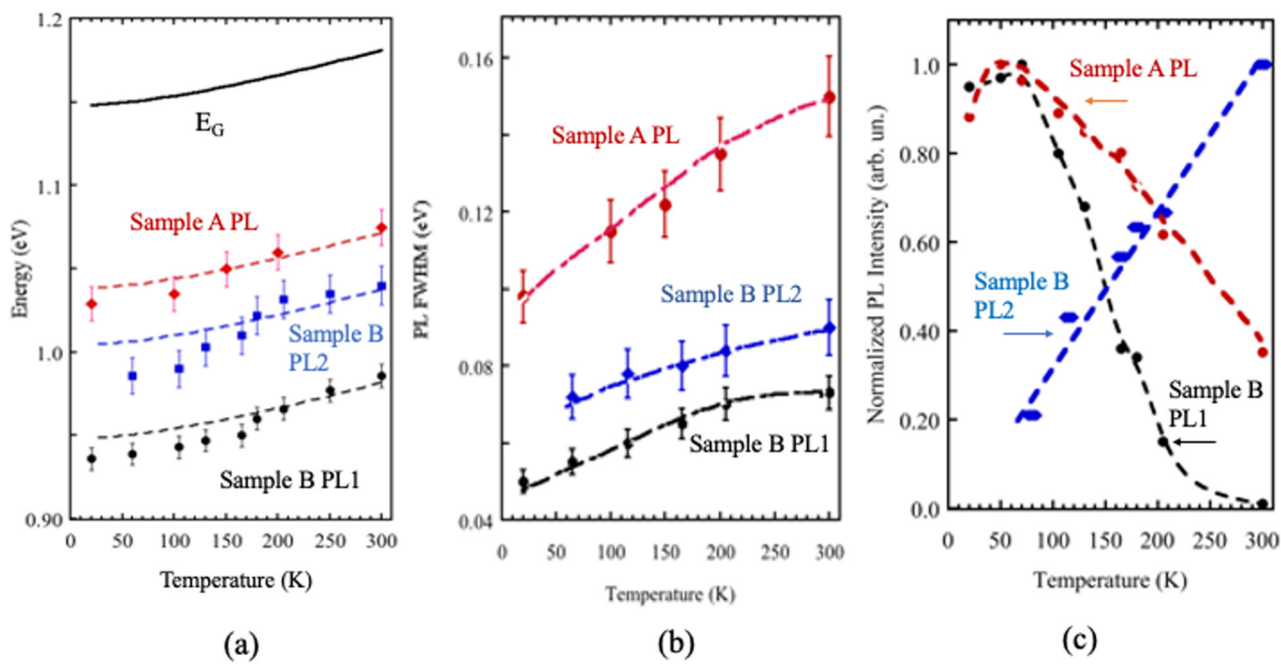

**FIG. 4.** Temperature dependence of (a) the PL peak photon energy, (b) the PL FWHM and (c) the PL integrated intensity for the marked PL bands in samples A and B.

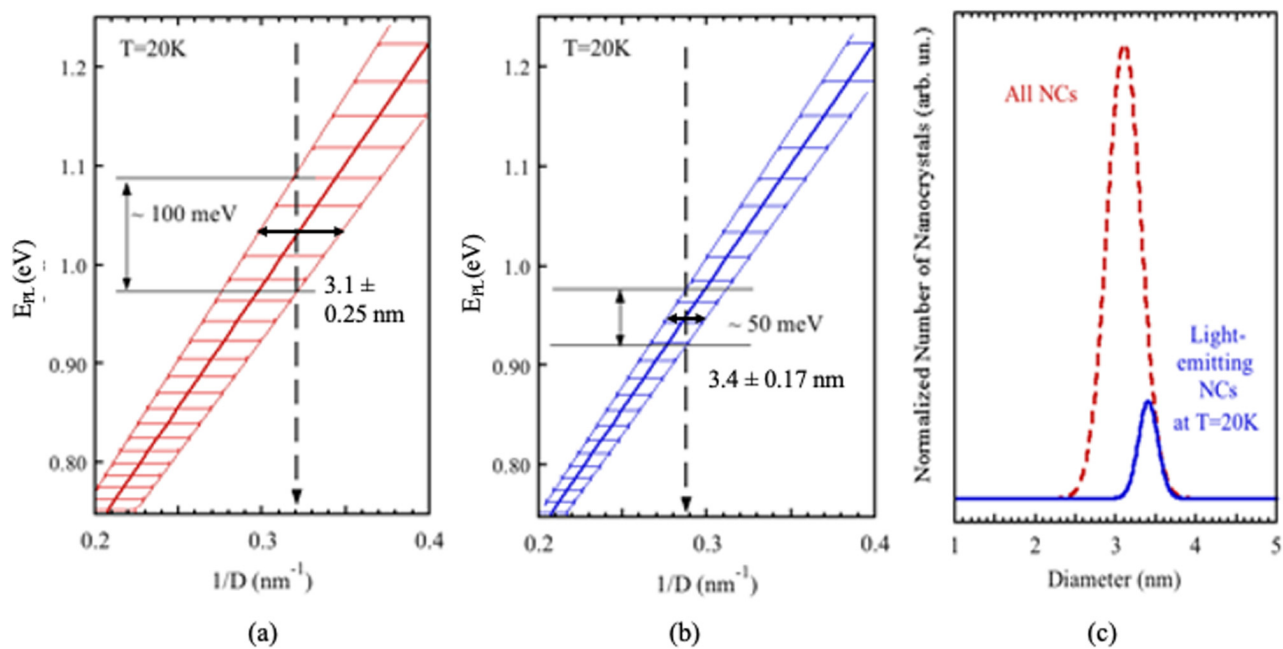

**FIG. 5.** Calculated low-temperature (T = 20 K) PL peak energies as a function of reciprocal diameters for PbS NCs with (a) 8% and (b) 5% diameter distributions (using Refs. 14, 34, and 35). Horizontal error bars show the NC size variations, and vertical arrows point on the PL peak energies in samples A and B. The energies of PL FWHM are shown for 3.1 ± 0.25 nm [Fig. 5(a)] and 3.4 ± 0.17 nm [Fig. 5(b)] PbS NC diameters. (c) Gaussian size distributions for (dashed line) PbS NC sizes observed in the TEM studies and (solid line) NC proposed to be responsible for light emission in sample B at T = 20 K.

In sample B, as the temperature decreases, the PL peak also shifts toward lower photon energies, and the PL FWHM decreases. However, the unexpected result is that at 100 K < T < 200 K, the PL spectrum clearly exhibits two (PL1 and PL2) bands [Fig. 3(b)]. Also, while the PL1 band intensity increases as the temperature decreases (similar to that in sample A), the PL2 band intensity exhibits the opposite behavior [Fig. 4(c)]. To explain these results, we assume the following:

(a) With an average distance between the PbS NCs of 1–2 nm, FRET is expected to be very efficient (Fig. 6).
(b) At room temperature, both donor-to-acceptor and back (from acceptor-to-donor) ET processes are possible [Fig. 6(a)]. Thus, there is no preferential ET, and, similar to that in sample A, the room temperature PL spectrum reflects the PbS NC size distribution and thermal broadening.
(c) As the temperature decreases, narrowing of the emission and absorption spectra due to reduced exciton–phonon interaction reduces the donor–acceptor spectral overlap, and that makes FRET more size selective [Fig. 6(b)]. Thus, the PL spectral features become more prominent, and longer- and shorter-wavelength PL bands (PL1 and PL2) can be separated [Fig. 3(b)]. Two different scenarios are considered. If back ET and exciton recombination at the donor NC site (which has a higher recombination rate in smaller size NCs[34]) is still possible, this process (which according to our data requires T > 60 K) is responsible for the shorter-wavelength PL component, or the PL2 band (Figs. 3 and 4). If back ET is completely suppressed and cascaded, multi-step FRET from smaller-to-larger NCs becomes the dominant process, the PL2 band disappears, and only the longer-wavelength PL component (the PL1 band) is observed [Fig. 6(b)]. Since the FRET characteristic time is expected to be considerably shorter compared to the exciton radiative lifetime in PbS NCs,[20–22,25] low-temperature cascaded FRET could be an efficient process.

As temperature decreases, the PL1 peak becomes significantly narrower and shifts toward lower photon energy faster than is expected due to Varshni's equation [Fig. 4(a)] due to the described low-temperature cascaded process with preferential ET to larger size NCs within the NC diameter distribution [Fig. 6(b)]. Thus, the Stokes shift increases as temperature decreases, similar to the data reported in Refs. 37 and 38. In agreement with our consideration, in sample B at T = 20 K, the PL peak photon energy is red shifted by ∼90 meV compared to that in sample A, and the PL peak FWHM is only 50 meV [Figs. 3(c) and 4(b)]. Using the calculated dependence of the energy gap as a function of reciprocal PbS NC diameter (based on Refs. 14, 34, and 35) and our experimental data, we conclude that at low temperature in films of densely packed PbS NCs with a 3.1 ± 0.25 nm diameter, most of the PL

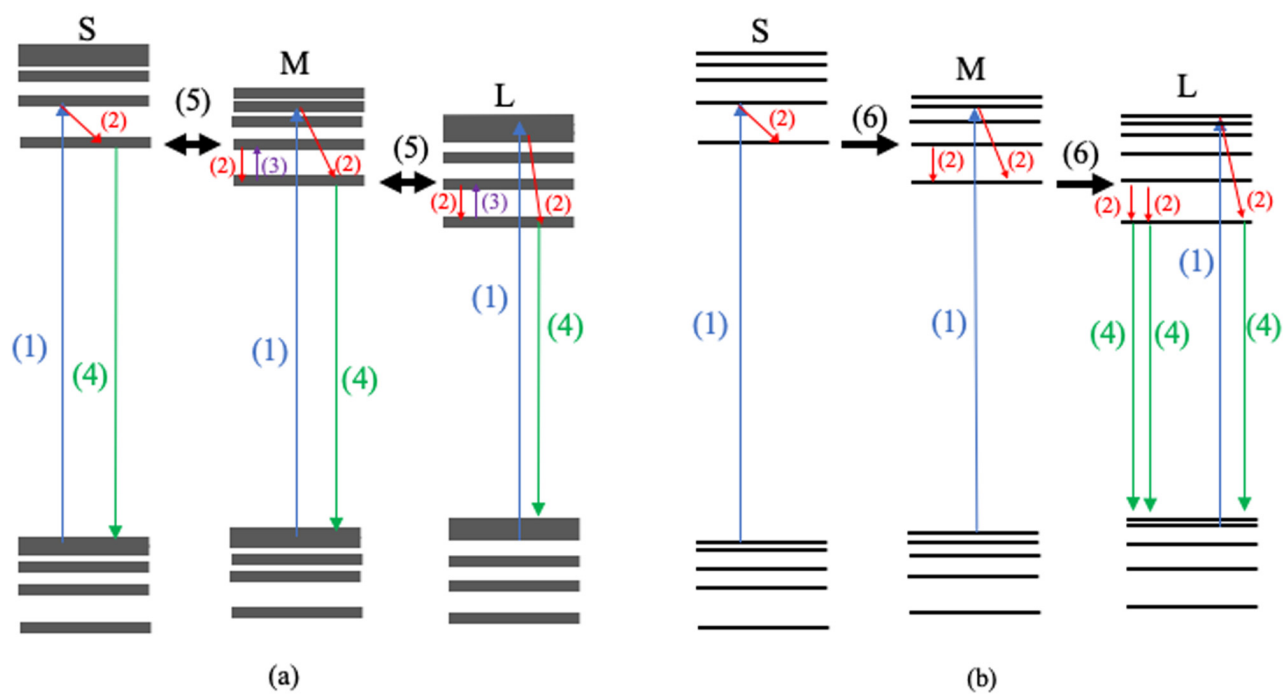

FIG. 6. Schematics of electronic processes in PbS NCs located within the FRET zone with three different sizes within the size distribution [described as small (S), medium (M), and large (L)]. The processes are shown at (a) room and (b) low temperatures and include (1) optical absorption, (2) electron thermalization, (3) thermal excitation, (4) exciton radiative recombination, (5) energy transfer with back transfer, and (6) energy transfer without back transfer. Note the narrowing of the energy levels at low temperature.

comes from NCs with a diameter of 3.4 ± 0.17 nm [marked as "light-emitting" NCs in Fig. 5(c)].

Based on our TEM data, sample B could be characterized as high-density PbS NC three-dimensional (3D) films, where NCs are separated by approximately 1–2 nm. This distance is comparable with a length of the organic ligands.[13,26] Our experimental results cannot categorically point out, which specific ET mechanism (e.g., FRET, DET, or direct electron–hole tunneling) is responsible for the observed PL properties in sample B. At the same time, FRET radius is estimated to be ∼5 nm,[22,24,25] DET zone is close to 1 nm,[25] the energy barrier between the NCs is estimated to be 1 eV or more,[36] and the tunneling electron is expected to have a heavy effective mass due to coupling to high-frequency vibrational modes of organic molecules at the NC surfaces.[39] Therefore, we consider FRET as the most probable mechanism of ET in sample B.

Compared to sample B, a quite different NC film morphology is found in samples prepared using solution with PbS NC concentration 10–15 mg/ml (sample C). In these samples, we consistently find two-dimensional (2D) clusters with a chain-like aggregation, each typically containing not more than 10 NCs (Fig. 7, inset). The cascaded ET mechanism is expected to be less efficient compared to that in densely packed, 3D NC films (sample B) but more probable compared to low-density NC monolayers with an average distance between NCs greater than the FRET radius (sample A). Thus, it is reasonable to expect that the low-temperature PL spectrum in sample C will be broader than 50 meV (sample B) but narrower than 100 meV (sample A). Also, the PL peak should be in between 0.93 eV (sample B) and 1.03 eV (sample A). In agreement with our expectations, we find that in sample C at T = 20 K, the PL peak is at 0.94 eV and the PL spectrum FWHM is ∼76 nm (Fig. 7).

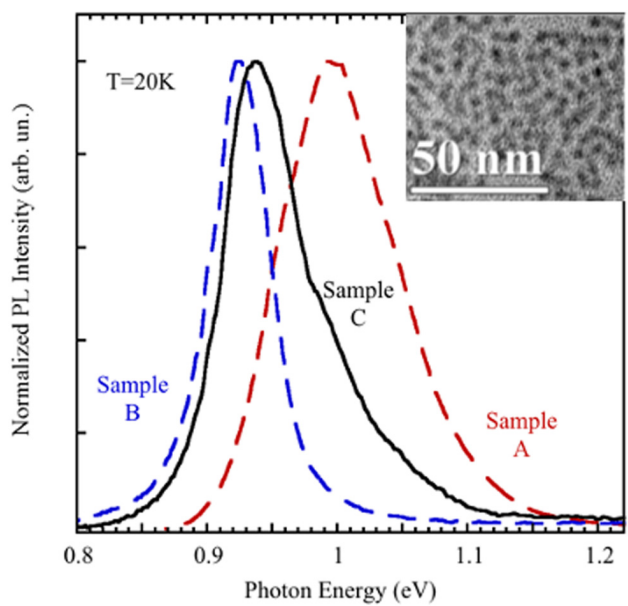

FIG. 7. Normalized low-temperature PL spectra in samples A, B, and C. The inset shows a TEM micrograph of a sample prepared using solution with NC concentration 10–15 mg/ml (similar to that in sample C) with clearly visible clusters of PbS NCs.

## V. CONCLUSION

In conclusion, our results point out that in PbS NC films, the PL properties, in addition to other factors, are controlled by the NC density and film morphology. At room temperature, in samples with high NC density, a combination of donor-to-acceptor and back ET processes provides a uniform distribution of excitons over the entire range of the available NC sizes. In samples with low NC density, where the ET process is suppressed, excitons are uniformly distributed as well. Thus, no significant difference is observed in the PL spectra in high- and low-density PbS NC film samples at room temperature.

In contrast, low-temperature PL spectra in the same samples are dramatically different. In low NC density samples, the PL spectrum entirely depends on the PbS NC size distribution. In high NC density samples, because back ET is suppressed, the dominant ET mechanism is cascaded, from smaller-to-larger NC diameter (within the NC size distribution) FRET. Thus, exciton recombination mostly takes place at larger size NCs within the NC size distribution, and it explains the observed PL peak narrowing (significantly below the value expected due to the average NC size) and strong red shift of the PL spectrum (also well below the photon energy associated with the average NC size). It is estimated that at low temperature in a sample of densely packed $3.1 \pm 0.25$ nm diameter PbS NCs, the PL comes from the NCs with $3.4 \pm 0.17$ nm diameter. In addition, it has been shown that film morphology, more specifically NC clustering, is also affecting the low-temperature PL spectra in PbS NC thin films. The presented results and conclusions suggest that similar effects can be found in other NC systems, where FRET is expected to play a similar role. Also, the study of time-resolved PL in NC films with different morphologies could be an interesting extension of this work.

## ACKNOWLEDGMENTS

This work is supported in part by the Semiconductor Research Corporation and the Foundation at NJIT. S. B. Hafiz gratefully acknowledges his graduate student support from the National Science Foundation (No. ECCS-1809112). The authors thank Alexander L. Efros and C. Delerue for discussions and Joan Mahon for proofreading the manuscript.

## DATA AVAILABILITY

The data that support the findings of this study are available from the corresponding author upon reasonable request.